\documentclass[aps,prl,reprint,superscriptaddress,showpacs]{revtex4-1}
\usepackage{color}
\usepackage{graphicx}
\usepackage{amsmath}
\usepackage{amssymb}
\usepackage{bm}
\usepackage{braket}
\usepackage{soul}
\usepackage{gensymb}
\graphicspath{{pict/}{Figs/}{}}

\begin{document}

\title{Two-photon tomography using on-chip quantum walks}
\author{James~G.~Titchener}
\author{Alexander~S.~Solntsev}
\author{Andrey~A.~Sukhorukov}

\affiliation{Nonlinear Physics Centre, Research School of Physics and Engineering, Australian National University, Canberra, Australian Capital Territory  2601, Australia}

\begin{abstract}

We present a conceptual approach to quantum tomography based on first expanding a
quantum state across extra degrees of freedom and then exploiting the
introduced sparsity to perform reconstruction. We formulate its
application to photonic circuits, and show that measured spatial
photon correlations at the output of a specially tailored
discrete-continuous
quantum-walk can enable full reconstruction of any two-photon
spatially entangled and mixed state at the input. This approach does
not require any tunable elements, so is well suited for integration
with on-chip superconducting photon detectors.

%
%
\end{abstract}

\maketitle

A key benefit of quantum systems is that their information capacity scales exponentially with the system size. However this also presents a challenge, as fully characterizing such a system requires a correspondingly large number of measurements.  Many measurements are required for two reasons. Firstly, the result of a single quantum measurement is probabilistic, so to get information about the underlying density matrix, the same measurement must be repeated many times on near identical states to calculate expectation values \cite{Fano:1957-74:RMP}. Secondly, the density matrix exists in a high dimensional space that cannot be fully accessed from one set of measured expectation values. Thus to characterize the density matrix a variety of linear transformations are usually applied to the system to measure many different expectation values.
Generally to characterize a set of $N$ qubits the measurement system is reconfigured to realize $4^N$ different linear transformations, and expectation values are measured after each~\cite{James:2001-52312:PRA}. For example the density matrix of a pair of polarization entangled photons can be found using 16 different measurement settings realized by rotating a series of wave-plates and polarizers \cite{White:1999-3103:PRL}.

In this Letter we show that reconfiguring the measurement setup is actually not necessary for quantum state tomography. We demonstrate that one specially chosen static measurement setting is sufficient to fully characterize unknown density matrices.
This insight means that the usually complex and error sensitive tomography process is simplified and noise introduced when reconfiguring the measurement setup is avoided.
This can be applied to make quantum state tomography more accessible in many different systems, from trapped atoms \cite{Karski:2009-174:SCI} to quantum states of light \cite{Shadbolt:2012-45:NPHOT}.  Furthermore, this could facilitate quantum tomography in situations where it was previously too complex to be practical.


Our approach is inspired by compressed sensing.
Conventionally in compressed sensing it is assumed that a signal is sparse in some basis, and this knowledge allows reconstruction of the signal from fewer measurements than suggested by the Nyquist-Shannon sampling theorem~\cite{Donoho:2006-1289:ITIT}.
This can be exploited for sub-wavelength optical imaging~\cite{Gazit:2009-23920:OE, Szameit:2012-455:NMAT} and could be applied for fast tomography of near pure quantum states~\cite{Gross:2010-150401:PRL} and quantum process tomography~\cite{Baldwin:2014-12110:PRA}. It has also been shown that knowing a quantum state is sparse can facilitate characterization of three-photon quantum states just by measuring two-photon coincidences~\cite{Oren:1411.2238:ARXIV}.

\begin{figure}[htbp]
		\centering
		\includegraphics[width=\columnwidth]{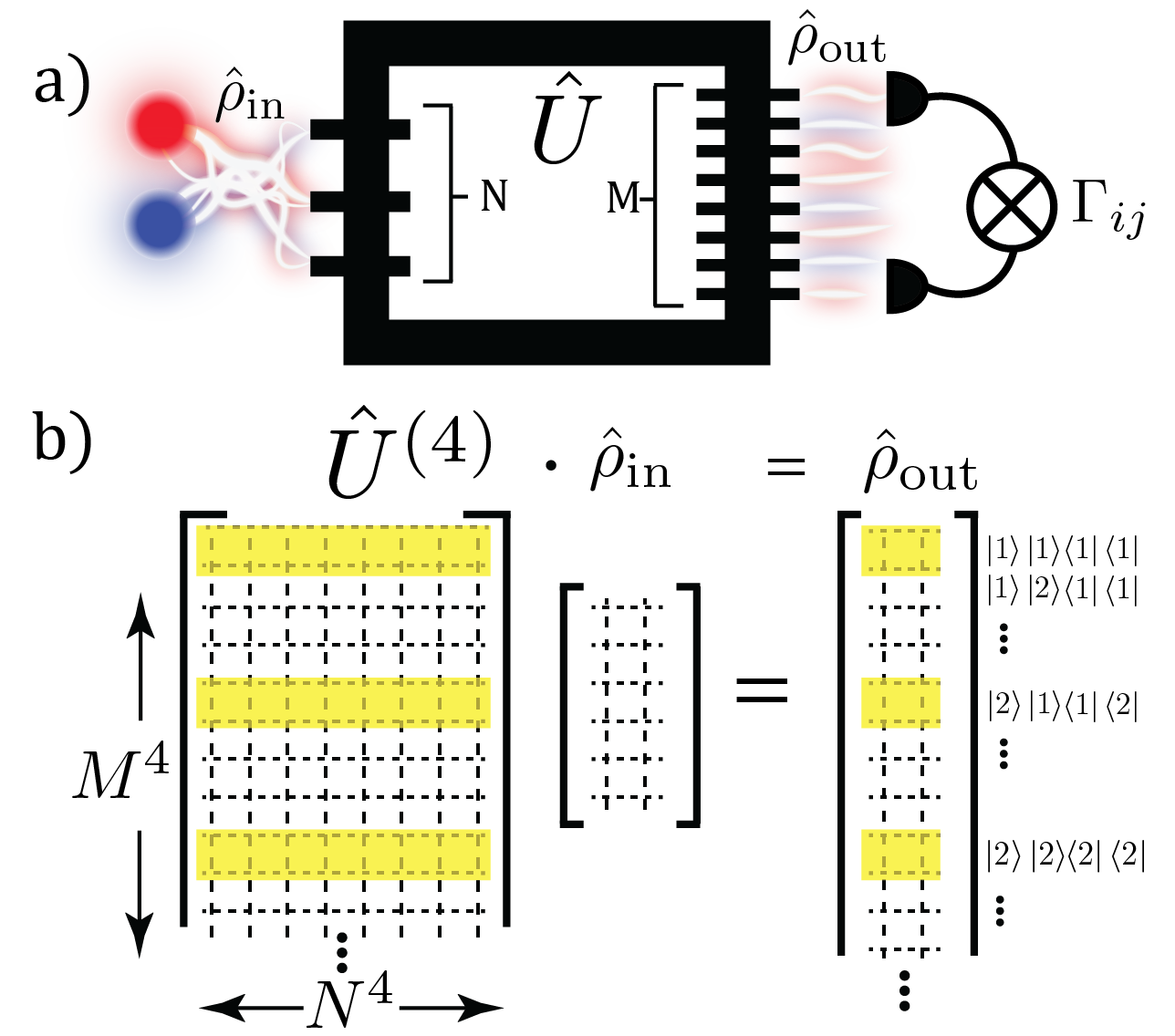}
		\caption{(a) A linear optical transformation operates on a mixed two photon state, then correlations between the photons are measured. (b) Linear mapping from input to output density matrix under the unitary transform. Highlighted rows and elements show the parts of the transformation associated with correlation measurements.}
		\label{fig:1}
	\end{figure}

However the assumption that a system is sparse is not valid for arbitrary quantum systems. Our key suggestion is to first force the unknown system to become sparse, and then apply a compressed sensing-like approach to tomography.
We achieve this by applying a linear transformation to the system that maps it to an increased number of modes [Fig.~\ref{fig:1}(a)], thus turning it into a sparse system. Careful choice of this transformation allows imaging of the full complex valued density matrix just from measuring one set of expectation values after the transformation. Thus the mixed quantum state can be fully characterized with a static measurement setup.

Here we illustrate our method by focusing on performing tomography of spatially entangled photon pairs. The transformation we will use to introduce sparsity is a hybrid of discrete and continuous quantum-walks.
In the past quantum-walks have been realized using atoms~\cite{Karski:2009-174:SCI}, ions~\cite{Schmitz:2009-90504:PRL} and pairs of photons~\cite{Peruzzo:2010-1500:SCI}. They have a wide range of potential applications, from helping to explain energy transfer in photosynthesis~\cite{Engel:2007-782:NAT} to providing a platform for universal quantum computation~\cite{Childs:2009-180501:PRL} and solving the boson sampling problem~\cite{Broome:2013-794:SCI, Tillmann:2013-540:NPHOT}.

Specifically we consider hybrid quantum-walks within on-chip coupled waveguide arrays (WGA)~\cite{Peruzzo:2010-1500:SCI}.
It has previously been was shown that a WGA can be used to perform interferometry~\cite{Minardi:2010-3009:OL, Minardi:2012-3030:OL, Shechtman:2013-24015:OE}, a classical analogue of quantum state tomography. Light can be coupled into a small number of waveguides, and the intensity measured at the output of the whole WGA. This allows both the phase and amplitude of the input classical fields to be determined, just from intensity measurements. The method also leverages sparsity (specifically, the knowledge that the input state was coupled into only a few selected waveguides) for full reconstruction of a complex valued input field from only intensity measurements.

We develop the approach to allow reconstruction of the two-photon density matrix from measurements of two-photon correlations, employing  %
a new class of WGA circuit realizing a hybrid of discrete and continuous quantum walks. These hybrid quantum walks map any input state entangled across $N$ ports to a sparse state at the output $M$ ports, where $M>N$ [Fig.~\ref{fig:1}(a)]. This approach is distinct from conventional compressed sensing because, instead of assuming that the signal is sparse, we introduce sparsity to the signal in a controlled way using the hybrid quantum-walk. Therefore the technique is not limited to imaging sparse inputs, it is capable of reconstructing any mixed quantum state and thus forms a complete system for quantum state tomography.

The ability to perform quantum state tomography in a static and 1-D quantum walk circuit is particularly promising when considered in the context of on-chip quantum photonics. Using normal tomography in an integrated photonic chip requires thermally tunable phase shifters to be reconfigured to perform different measurements~\cite{Shadbolt:2012-45:NPHOT, Carolan:2015-711:SCI}. The latest technological advances enable on-chip integration of highly efficient superconducting photon detectors~\cite{Najafi:2015-5873:NCOM}, however integration of multiple tunable phase shifters operating at low temperature remains an open problem. Our approach to tomography would have the advantage of being easily integrated with on-chip detection schemes since it requires no tunability.

To demonstrate our approach to tomography, we first introduce the mathematical formalism to propagate a mixed quantum state through a quantum walk and take two-photon correlation measurements at the output.
We consider a mixed state comprised of two photons propagating through a waveguide circuit. The net effect of the quantum walk on a single photon can be written as a linear optical transformation, $\boldsymbol{\hat{U}}$ \cite{Carolan:2015-711:SCI}. This transformation will operate on the wavefunction of each photon coupled into the circuit according to
	$\ket{\psi}^{\text{out}} = \boldsymbol{\hat{U}} \ket{\psi}^{\text{in}}$.
%
	For simplicity we assume that both photons undergo the same unitary transformation, but the following could be easily generalized to the case where each photon undergoes a different transformation.

	The transformation operating on the joint wavefunction of the two photons ($\Psi=\sum_i c_i \ket{\psi_1}^{(i)}\ket{\psi_2}^{(i)}$) will read
	$\Psi^{\text{out}} = \boldsymbol{\hat{U}}\otimes \boldsymbol{\hat{U}}  ~ \Psi^{\text{in}}$.	
	Now a statistical mixture of two photon wavefunctions is best described by the density matrix, $\hat{\rho} = \sum_j p_j \ket{\Psi_j} \bra{\Psi_j}$, where $p_j$ is the probability of the mixed quantum system being in state $\ket{\Psi_j}$ \cite{Fano:1957-74:RMP}. 
We see that under the transformation $\boldsymbol{\hat{U}}$ the density matrix changes as
	\begin{equation}
	\hat{\rho}_{\text{out}} = \boldsymbol{\hat{U}}\otimes \boldsymbol{\hat{U}}  ~ \hat{\rho}_{\text{in}} ~ \boldsymbol{\hat{U}}\otimes \boldsymbol{\hat{U}} =  \boldsymbol{\hat{U}}^{(4)}  \hat{\rho}_{\text{in}},
	\label{linear_map}
	\end{equation}
		where $ \boldsymbol{\hat{U}}^{(4)}  = \boldsymbol{\hat{U}}\otimes \boldsymbol{\hat{U}}\otimes\boldsymbol{\hat{U}}^{*}\otimes \boldsymbol{\hat{U}}^{*}$. Eq.~(\ref{linear_map}) can be expressed as a matrix equation as shown in Fig.~\ref{fig:1}(b). Here the input and output density matrices are written as vectors of $N^4$ and $M^4$ elements respectively, $N$ being the number of input waveguides and $M$ the number of output waveguides. The transformation $\boldsymbol{\hat{U}}^{(4)}$ can then be rewritten as a $M^4 \times N^4$ matrix in a consistent way.
	
At this point it is important to consider how the output state will be measured. Typically in integrated quantum photonics the measured quantity will be the correlations in the arrival time of two photons at any of the output waveguides \cite{Shadbolt:2012-45:NPHOT}. This will give the probability amplitudes, $\Gamma_{ij}$, associated with observing one photon in waveguide $i$ and the other in waveguide $j$. Mathematically, expectation values of these probability amplitudes are represented by applying the measurement operator $\ket{j}\ket{i}\bra{i}\bra{j}$ to the density matrix \cite{Fano:1957-74:RMP}, and accordingly
%
	$\Gamma_{ij} = \bra{i}\bra{j}\hat{\rho}_\text{out}\ket{j}\ket{i}$.
%
These amplitudes correspond to the measurement of only a sub-set of the whole output density matrix. We illustrate this schematically in Fig.~\ref{fig:1}(b), where such measurable elements are highlighted in yellow.
%
Our goal is to use these measurements to reconstruct all the elements of the input density matrix.
Accordingly, we reformulate Eq.~(\ref{linear_map}) by excluding the unobservable elements of $\hat{\rho}_{\text{out}}$,
\begin{equation} \label{(inverse)}
	\Gamma=  \boldsymbol{M}^{\Gamma}~ \hat{\rho}_{\text{in}} .
\end{equation}
Here $\boldsymbol{M}^{\Gamma}$ is a matrix containing only the rows of $\boldsymbol{\hat{U}}^{(4)}$ that map the input state ($\hat{\rho}_\text{in}$) to correlations ($\Gamma$) in the output mixed state.
	
Now the problem of quantum state tomography can be formulated as follows; can the input density matrix, $\hat{\rho}_{\text{in}}$, be inferred from Eq.~(\ref{(inverse)}) given that $\Gamma$ has been measured? To answer this question we first consider the dimensionality of $\hat{\rho}_{\text{in}}$ and $\Gamma$. There are generally $M^2$ different correlations which can be measured, each being a real number. 
The input two-photon density matrix contains $N^4$ elements, most of which are complex. However since it is Hermitian, it is fully defined by $N^4$ real parameters. For the tomography problem to be solvable, the amount of measured information should be equal to or exceed the number of unknowns, and accordingly we require $M^2 \ge N^4$. Furthermore, if the two photons are indistinguishable, which we consider in the examples below, due to additional symmetries the condition reduces to
\begin{equation} \label{MNcondition}
	\frac{M (M+1)}{2} \ge \frac{N^2 (N+1)^2}{4}.
\end{equation}
%
%
This dimensionality requirement means that the transformation $\boldsymbol{\hat{U}}$ must be a sparse one, mapping $N$ inputs to $M$ outputs with 
$M>N$.
If the requirement is satisfied, then it is possible that Eq.~(\ref{(inverse)}) can be inverted to find the input density matrix. The inversion can be carried out using the Moore-Penrose pseudoinverse~\cite{Press:2007:NumericalRecipes},	
	\begin{equation}
	\hat{\rho}_{\text{in}} = (\boldsymbol{M}^{\Gamma})^{-1} \Gamma .
	\label{(inverse2)}
	\end{equation}
Here the pseudoinverse is defined as $(\boldsymbol{M}^{\Gamma})^{-1} = V S^{-1} U^{\dagger}$, where a singular value decomposition is performed as $\boldsymbol{M}^{\Gamma} = U S V^{\dagger}$.
If Eq.~(\ref{(inverse2)}) maps every possible set of correlations to a unique density matrix, then full quantum state tomography can be achieved simply by knowing the correlations, $\Gamma$, and $(\boldsymbol{M}^{\Gamma})^{-1}$.
Of course, $(\boldsymbol{M}^{\Gamma})^{-1}$, the mapping back from measured correlations to input density matrix, is completely determined by $\boldsymbol{\hat{U}}$, the linear optical transform implemented on each photon before correlation measurement. This mapping must be carefully chosen so that Eq.~(\ref{(inverse2)}) produces a unique solution for any measured correlations, making quantum state tomography possible.

\begin{figure}[htbp]
		\centering
		\includegraphics[width=\columnwidth]{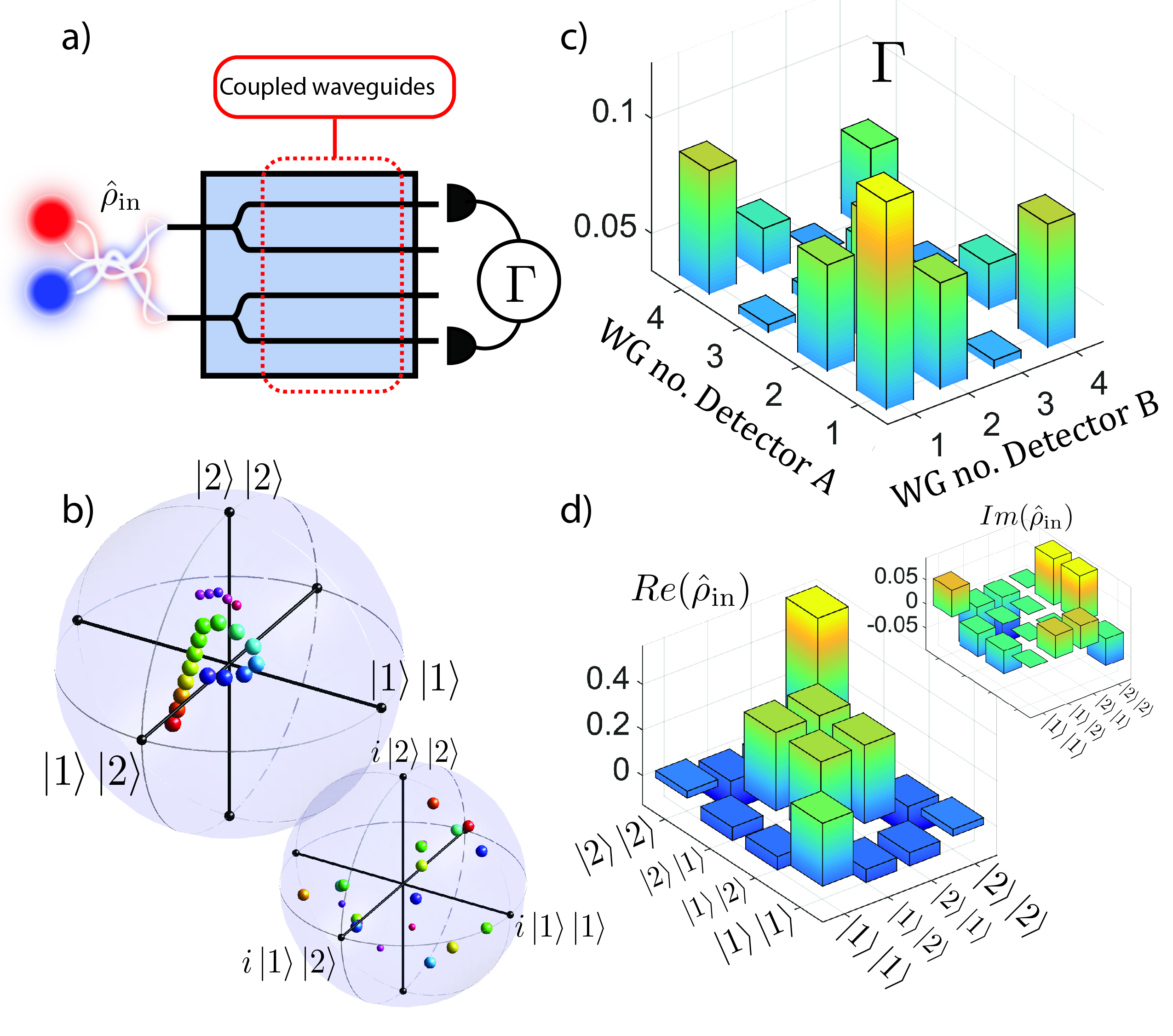}
		\caption{(a) Diagram of a hybrid quantum walk for two-photon tomography. 
(b) Real and imaginary parts of the input mixed state. Each wavefunction in the mixture is represented by a different colored sphere, statistical weight of each wavefunction is shown by the diameter of the sphere.
(c) Simulated correlation measurements at the output.  
(d) Real and imaginary parts of the input density matrix, recovered using correlation measurements shown in~(c).}
		\label{fig:2}
\end{figure}

We observe that continuous quantum walks in planar WGA's do not lead to unique solutions of Eq.~(\ref{(inverse2)}), and thus cannot be used for tomography of mixed quantum states. Previous works with WGA's have been focused on recovering classical amplitude profiles and mutual coherence functions of classical light fields~\cite{Shechtman:2013-24015:OE, Minardi:2015-13804:PRA} or the density matrix of
quantum states where the form of the entangled part of the wavefunction is known a priori~\cite{Oren:1411.2238:ARXIV}. Furthermore it has been shown that recovery of the mutual coherence function requires nonlocal coupling in the WGA \cite{Minardi:2015-13804:PRA}, which can be achieved in a 2-D WGA. This result extends to quantum state tomography also due to the similarities between incoherent classical light and mixed quantum states. The use of a 2-D WGA may be undesirable, especially for the goal of fully integrated quantum photonics on a planar chip. Thus we develop a special type of optical circuit, combining a discrete time quantum walk with a continuous quantum walk. As we demonstrate in the following, such a hybrid quantum walk can allow full tomography of a two-photon state in a planar device.

An example of a linear optical circuit implementing a suitable hybrid quantum walk is shown in Fig.~\ref{fig:2}(a). Here two input waveguides ($N=2$) are split into an array of four, realizing one step of a discrete quantum walk. Following this the four waveguides form a coupled waveguide array ($M=4$), with dimensionless length $L=3.76$ and coupling rate $C=1$. We note that the waveguide numbers satisfy the necessary condition in Eq.~(\ref{MNcondition}). This means the circuit has the potential to allow quantum state tomography just by taking correlation measurements at the output.

We use this circuit to demonstrate tomography of a mixed state comprised of 20 different pure states as shown in Fig.~\ref{fig:2}(b). This way of representing the mixed state is used to demonstrate that the state is nontrivial, containing many entangled two-photon pure states.
This representation of a mixed state was then used to calculate the input density matrix, $\hat{\rho}_\text{in}$, which is the typical (and most efficient) representation of such a state.
We then propagate the density matrix through the hybrid WGA circuit, and model the output two-photon spatial correlations presented in Fig.~\ref{fig:2}(c).
From the correlation measurements the input density matrix is uniquely recovered using Eq.~(\ref{(inverse2)}). The real and imaginary parts of the recovered input density matrix are shown in Fig.~\ref{fig:2}(d), and we have verified that they exactly match the input state.

	\begin{figure}[htbp]
		\centering
		\includegraphics[width=\columnwidth]{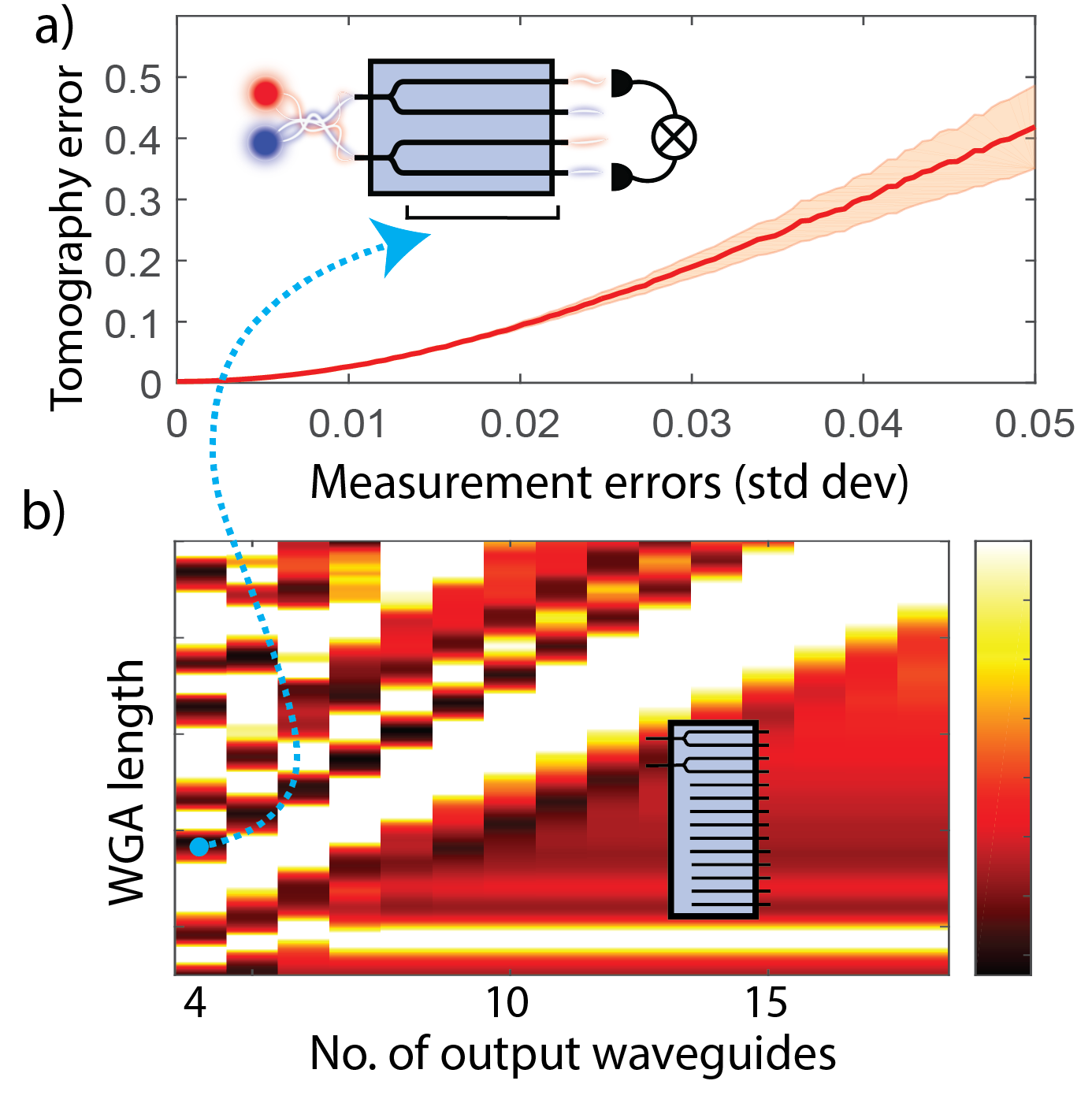}
		\caption{(a) Mean error in the tomographic reconstruction of randomly generated mixed states using the optical circuit shown. The standard deviation of simulated Gaussian error in the correlation measurements is plotted along the x-axis. (b) The condition number of different hybrid quantum walking circuits. The x-axis shows increasing number of waveguides in the WGA, while the y-axis shows increasing length of the array. }
		\label{fig:3}
	\end{figure}
Any experimental implementations of tomography will suffer from errors in the detection of photons, thus the tomographic technique must be robust.
In Fig.~\ref{fig:3}(a) we show the error tolerance of the optical circuit from Fig. \ref{fig:2}(a). Gaussian error is added to the correlation measurements, then tomography of a random mixed state is attempted. The x-axis shows increasing standard deviation of the Gaussian noise, while the y-axis shows the mean least squares error in the tomographic reconstruction, each point averaged over 10000 attempts at reconstructing random density matrices. This shows that robust tomography of a two-photon state is possible in the presence of errors.
	
More generally the stability of the technique is determined by the condition number~\cite{Press:2007:NumericalRecipes} of the transformation $\boldsymbol{M}^{\Gamma}$. A high condition number means inversion will be impossible in realistic situations, because errors in the measurements will be highly amplified in the recovered density matrix. In Fig.~\ref{fig:3}(b) the condition number of the transformation is plotted against the number of output waveguides and the length of the WGA. We see that for the case of only four output waveguides discussed above optimal choice of the length can give a sufficiently low condition number to allow tomography. Increasing the number of output waveguides can further reduce the condition number of the transformation. Although with more waveguides more single photon detectors are required, and the complexity of the device increases.
	
	\begin{figure}[htbp]
		\centering
		\includegraphics[width=\columnwidth]{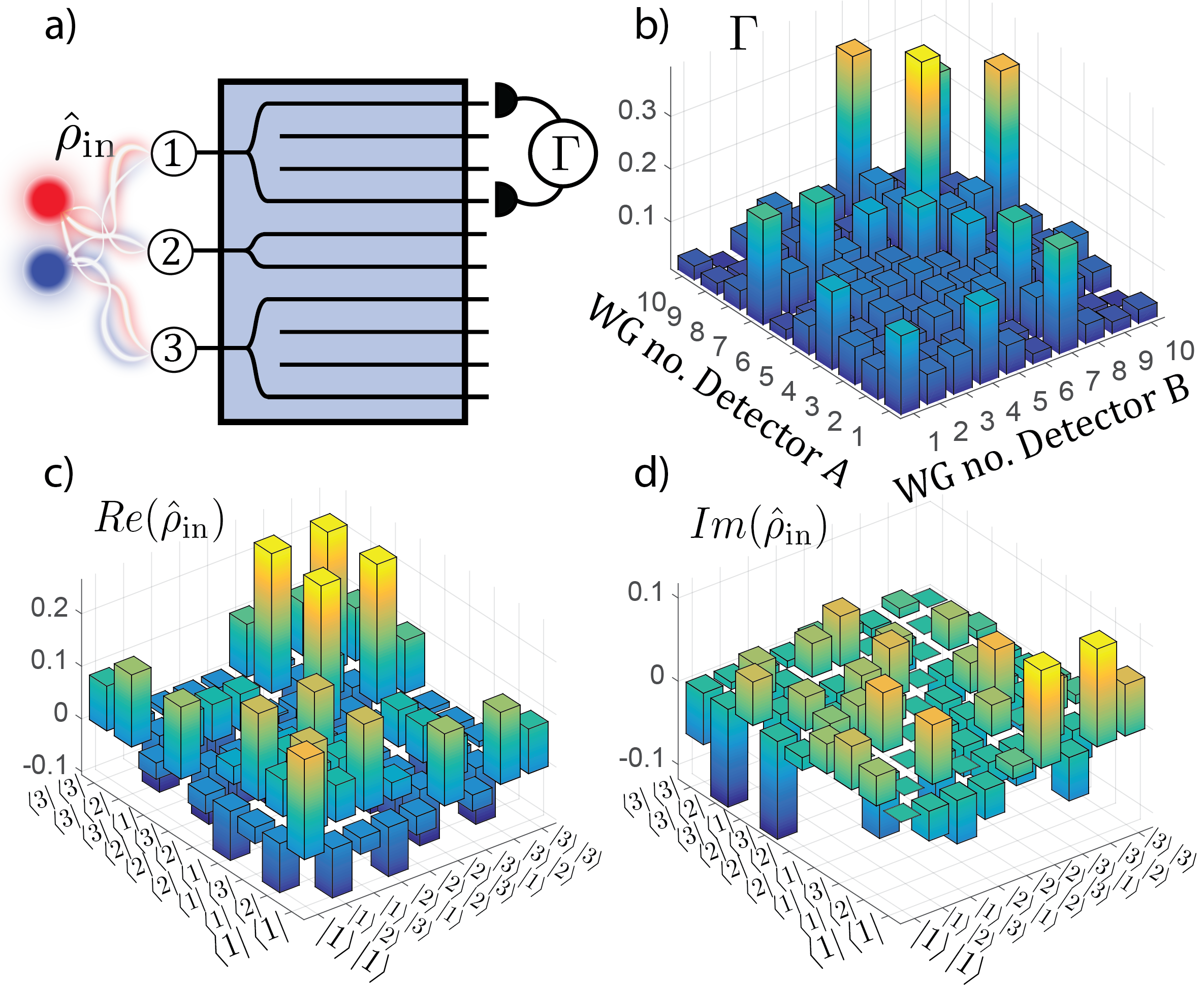}
		\caption{(a)~Diagram of linear optical circuit for tomography of a two-photon mixed state coupled into three input waveguides. (b)~Simulated output correlation measurements. (c)~Real and (d)~imaginary parts of the input density matrix reconstructed from correlation measurements.}
		\label{fig:4}
	\end{figure}

Tomography of states with more input degrees of freedom is also possible. As an example, in Fig.~\ref{fig:4}(a) we demonstrate a static optical circuit, which can be used to recover the density matrix of any two-photon mixed state coupled into three input waveguides. The circuit is again a hybrid quantum walk, splitting three input waveguides into six then allowing them to undergo a continuous quantum walk in a ten waveguide array. According to Eq.~(\ref{MNcondition}), eight output waveguides could provide enough information to be measured via correlations in order to determine the input density matrix. However we chose to use ten output waveguides to increase the robustness of the tomography to errors. The output correlations are shown in Fig.~\ref{fig:4}(b). The reconstructed real and imaginary parts of the input density matrix are presented in Figs.~\ref{fig:4}(c) and (d), respectively, and they exactly match the input state.

In conclusion, we have introduced a method of performing quantum state tomography that only requires a single set of expectation values to be measured in a static system, in contrast to usual approaches with reconfigurable elements.
We have developed hybrid quantum walk circuits for
spatially entangled photon-pair tomography, where the input state is reconstructed from output correlation measurements.
This could facilitate on-chip quantum state tomography with integrated superconducting single-photon detectors and no tunable elements.

There are a diverse variety of other systems that could also benefit from this type of tomography. They include  nuclear magnetic resonance qubits \cite{Ryan:2005-62317:PRA}, Bose-Einstein~\cite{Morsch:2006-179:RMP} and exciton polariton condensates~\cite{Byrnes:2014-803:NPHYS}, atoms~\cite{Karski:2009-174:SCI} and ions~\cite{Zahringer:2010-100503:PRL} in lattice potentials.
Quantum walks have been demonstrated in a number of these systems, making the correspondence to the two-photon tomography shown here quite direct. However quantum walks are not a fundamental requirement for this new approach to tomography. 
Our key concept is that a linear transformation can be applied to the system to introduce sparsity. Our approach shows how to design this transformation to map every possible density matrix to a unique set of measurable expectation values, thus enabling quantum state tomography.



\end{document}